\documentclass[aps,pre,reprint,superscriptaddress,longbibliography,showpacs]{revtex4-1}
\usepackage{amsmath,amssymb,graphicx,xcolor}
\usepackage{times,bm,url}
\usepackage[colorlinks=true,breaklinks=true,allcolors=blue]{hyperref}
\newcommand{\orcidicon}[1]{\href{https://orcid.org/#1}{\includegraphics[height=\fontcharht\font`\B]{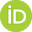}}}

\newcommand{\ex}[1]{\ensuremath{\langle #1 \rangle}} 

\DeclareMathOperator{\tr}{{Tr}}

\def\Oo{\ensuremath{{\cal O}}} 
\def\Jj{\ensuremath{{\cal J}}} 
\def\Ll{\ensuremath{{\cal L}}} 
\def\Hh{\ensuremath{{\cal H}}} 
\def\Nn{\ensuremath{{N}}}
\def\Cc{\ensuremath{{\cal C}}}

\def\Tt{\ensuremath{{\cal T}}} 
\def\Gg{\ensuremath{{\cal G}}} 
\def\Qq{\ensuremath{{\cal Q}}}

\def\p{\ensuremath{{\partial}}}
\def\i{\ensuremath{{\imath}}}

\def\I{\ensuremath{{\alpha}}}

\newcommand{\nint}[3][]{\ensuremath{ \int_{\mbox{\scriptsize $#1$}}^{\mbox{\scriptsize $#2$}}\!\!\! \mbox{\small$\!\mathrm{d}#3$\! }}}
\makeatletter
\let\cat@comma@active\@empty
\makeatother

\makeatletter
\g@addto@macro\bfseries{\boldmath}
\makeatother

\begin{document}
\title{Thermalization of many many-body interacting SYK models}

\author{Jan C.\ Louw \!\orcidicon{0000-0002-5111-840X}}
\affiliation{Institute for Theoretical Physics, Georg-August-Universit{\"a}t G{\"o}ttingen,  Friedrich-Hund-Platz 1, 37077~G{\"o}ttingen, Germany}

\author{Stefan Kehrein}
\affiliation{Institute for Theoretical Physics, Georg-August-Universit{\"a}t G{\"o}ttingen, Friedrich-Hund-Platz 1, 37077~G{\"o}ttingen, Germany}

\date{\today}
\begin{abstract}
We investigate the non-equilibrium dynamics of complex Sachdev-Ye-Kitaev (SYK) models in the $q\rightarrow\infty$ limit, where $q/2$ denotes the order of the random Dirac fermion interaction. We extend previous results by Eberlein et al. \cite{Eberlein2017} to show that a single SYK $q\rightarrow\infty$ Hamiltonian for $t\geq 0$ is a perfect thermalizer in the sense that the local Green's function is instantaneously thermal. The only memories of the quantum state for $t<0$ are its charge density and its energy density at $t=0$. Our result is valid for all quantum states amenable to a~$1/q$-expansion, which are generated from an equilibrium SYK state in the asymptotic past and acted upon by an arbitrary combination of time-dependent SYK Hamiltonians for $t<0$. Importantly, this implies that a single SYK $q\rightarrow\infty$ Hamiltonian is a perfect thermalizer even for non-equilibrium states generated in this manner. 

\end{abstract}

\maketitle

\section{Introduction} 

The thermalization of closed quantum many-body systems has become a major research topic due to its relevance both for the foundations of quantum statistical mechanics \cite{Polkovnikov2011Aug,Ueda2020Dec} and for experiments, especially in cold atomic gases \cite{Kinoshita2006Apr}. Unitarity of time evolution in a closed system implies that a pure state can never evolve to a mixed state described by a thermal density operator. However, a time evolved pure state can become indistinguishable from a thermal state from the point of view of local measurements or, more generally, measurements of few-body operators. It is in this sense that thermalization of closed quantum many-body systems is usually defined \cite{Gogolin2016Apr}. 

The two main categories of thermalization behavior are that of integrable systems, that generically time evolve to a non-thermal stationary state described by a generalized Gibbs ensemble (GGE) \cite{Rigol2007Feb,Essler2016Jun,Vidmar2016Jun}, and non-integrable systems, whose stationary state can be described by a thermal state. The generic underlying fundamental reason for the thermalization of non-integrable systems is the eigenstate thermalization hypothesis (ETH) \cite{Deutsch1991Feb,Srednicki1994Aug,Rigol2008Apr,Deutsch2018Jul}. In between these two categories are strongly disordered systems, which can show many-body localization with non-ETH behavior for large disorder, and ETH behavior for weaker disorder \cite{Abanin2019May}. 

The actual thermalization dynamics for non-integrable systems is usually described by a quantum Boltzmann equation (QBE). However, the QBE is only applicable in systems that allow a quasiparticle description \cite{Kamenev2011}. Within the QBE framework, the approach to equilibrium is exponential with a relaxation time $\tau$ that scales like $1/\tau\sim U^2\,T^2$ at low temperatures, where $U$ is the interaction strength and $T$ is the temperature of the final state \cite{sachdev2001}.    

An important class of materials where the quasiparticle picture of Fermi liquid theory is invalid are strange metals based on their linear in $T$ electrical resistivity behavior \cite{Greene2020Mar,Legros2019Feb}. In the past few years, the Sachdev-Ye-Kitaev (SYK) model \cite{Sachdev1993May,Kitaev2015} has paved the way to a better understanding of such materials \cite{Sachdev2015,Song2017}. Apart from its lack of a quasiparticle description, the SYK model has other fascinating properties like being analytically solvable in the thermodynamic limit while at the same time being chaotic (even maximally chaotic in a well-defined sense at low temperatures \cite{Maldacena2016}) \cite{Maldacena2016Nov} and connections to holographic theories and black holes \cite{Sachdev2015,Kitaev2015}. The original SYK model contains only an interaction term for $q=4$ Majorana fermions \cite{Kitaev2015}, but generalizations to general $q$-particle interaction terms and even superpositions of different $q$-interaction terms are possible while still retaining the analytic solvability in equilibrium \cite{Maldacena2016Nov}. The same is true for Dirac fermions instead of Majorana fermions \cite{Sachdev2015}. Of particular interest is the many many-body limit $q\rightarrow\infty$, where calculations become analytically more manageable \cite{Maldacena2016Nov}. 

In this paper we are interested in the thermalization dynamics of the SYK model \cite{Eberlein2017,Sonner2017Nov,Magan2016Aug,Bhattacharya2019Jul,Bandyopadhyay2021Aug,Zanoci2021Sep,Cheipesh2021Sep,Almheiri2019Dec,Haldar2020Mar,Kuhlenkamp2020Mar,Larzul2021Jul}. Due to its lack of quasiparticles the relaxation time is expected to be 'Planckian', $1/\tau=f\,k_B\,T/\hbar$, for low temperatures, where $f$ is a constant of order~1 \cite{sachdev2001}. In the low-temperature limit the relaxation time is therefore both much shorter and universal with no dependence on the interaction strength as compared to e.g. Fermi liquid theory. The thermalization dynamics of the SYK model for Majorana fermions after a quench was first investigated by Eberlein et al. \cite{Eberlein2017}. They showed how the analytic solvability of the SYK model in equilibrium carries over to non-equilibrium situations, which can be described by a finite set of integro-differential equations.  These equations could then be solved numerically, or even analytically in the limit of $q\rightarrow\infty$ interacting Majorana fermions. Specifically, they presented numerical results for a quench starting from an equilibrium state generated by an SYK $q=2$ plus $q=4$ model to a $q=4$ model that are consistent with $1/\tau\propto T$. 

In the $q\rightarrow\infty$ limit Eberlein et al. could solve the Kadanoff-Baym equations analytically for a quench starting from an equilibrium state generated by an SYK $q$ plus $2q$ model (or alternatively: $q$ plus $q/2$ model). The post-quench Hamiltonian was a single SYK $q$~model. The surprising result was instantaneous equilibrium behavior of the local Green's function after the quench \cite{Eberlein2017}, implying that there is no memory of the pre-quench state except for its energy density. 

\begin{figure}[h]
	\centering
	\includegraphics[width=1\linewidth]{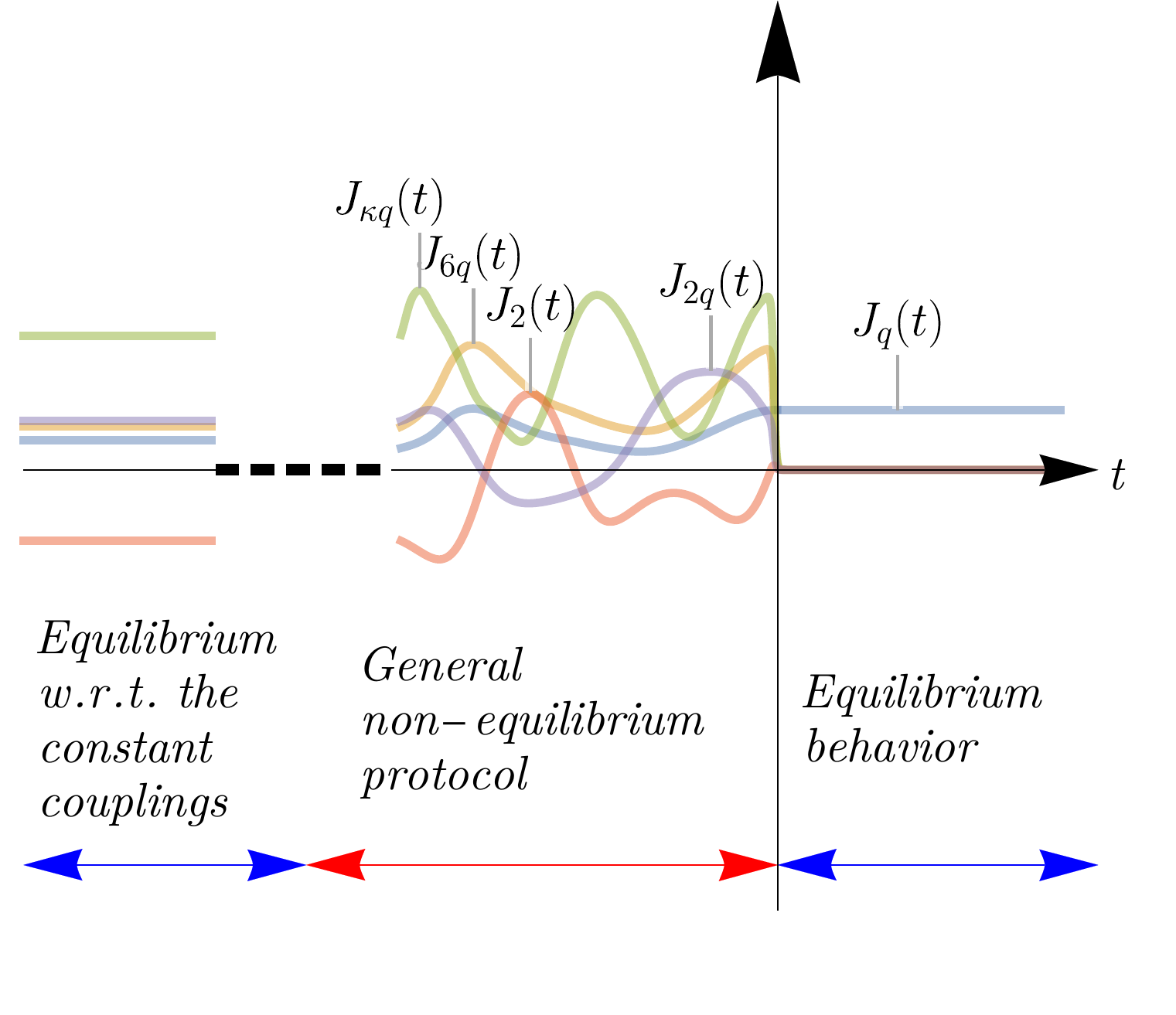}
	\caption{A schematic summary of the key result of this work. In the asymptotic past, the system is prepared in a thermal Gibbs state with respect to initially time-independent couplings, see \eqref{Jpdef}. Afterwards the state is time evolved under a general sum of SYK models. For times $t\ge 0$ only a single SYK term remains. Our key result is that the system is then instantaneously in equilibrium, in the large $q$ limit.}
	\label{fig:coupling}
\end{figure}

Our paper generalizes the large-$q$ results in Ref.~\cite{Eberlein2017} along various lines. First of all, our analytic calculation holds generally for Dirac fermions (notice that the half-filled Dirac fermion SYK model is equivalent to the Majorana SYK model). More importantly, the system does not need to be in equilibrium before the quench, but can be in a general non-equilibrium state generated by a superposition of arbitrary time-dependent SYK interaction terms. Finally, we do not require a quench but just some arbitrary time-dependent protocol, see fig. \ref{fig:coupling}, that leads to a single remaining SYK term for $t\geq 0$. In the asymptotic past, $t\to-\infty$, the system is in a thermal equilibrium Gibbs state. The protocol, with arbitrary time-dependent couplings, then leads to non-equilibrium (NEQ) physics, which prepare an NEQ initial state $\varrho(t=0)$ . Under these conditions we show that the local Green's function is instantaneously in equilibrium for $t\geq 0$. The only memories of the quantum state for $t< 0$ are its charge density and its energy density at $t=0$. In this sense the $q\rightarrow\infty$ SYK Hamiltonian is a perfect thermalizer. The key requirements of our analytic calculation are the $q\rightarrow\infty$ limit and the existence of a single SYK term for $t\geq 0$. 

\emph{Outline}: We start by describing the general model in Sec. \ref{Model}. In Sec. \ref{Dynamics}, we go on to study the non-equilibrium dynamics given by the Kadanoff-Baym (KB) equations. We make use of the particular expansion allowed in the many many-body regime, described in Sec. \ref{ss:Many}, which significantly simplifies the (KB) equations. 

The main focus of our study, namely studying the dynamics of a very general state under a single SYK $q\rightarrow\infty$ Hamiltonian, is presented in Sec. \ref{s:Single}. For this case we obtain exact results for the local Green's function, which are shown to instantaneously satisfy all conditions of a thermal state in Sec. \ref{sec:Eq}. The equilibrium properties of this state, such as the energy density, are elaborated on in Sec. \ref{sec:PropState}. For completeness, we discuss the simplest case where all interactions are switched off, leaving only a kinetic term in Sec. \ref{Kinetic}. Finally, in Sec. \ref{conclusion} we summarize the results.

\section{Model \label{Model}}

The complex $p/2-$body interacting Sachdev-Ye-Kitaev (SYK) model is defined by all-to-all interactions \cite{Fu2018, Sachdev2015,Gu2020Feb}
$$
\Hh_{p} = \sum_{\substack{1\le i_1< \cdots < i_{p/2}\le \Nn \\ 1\le j_1<\cdots< j_{p/2}\le\Nn}} X^{ i_{1}\cdots  i_{p/2}}_{j_{1}\cdots j_{p/2}} c^{\dag}_{ i_1} \cdots c^{\dag}_{ i_{p/2}} c_{j_{p/2}} \cdots c_{j_1}. 
$$
Here $c_k^\dag, c_k$ are fermionic creation and annihilation operators respectively, while $\Nn$ is the number of lattice sites. The couplings, $X$, are complex random variables with zero mean.Their variance is given by
$$\overline{\,|X^{ i_{1}\cdots i_{p/2}}_{j_{1} \cdots j_{p/2}}|^{2}} =  \frac{U_{p}^2 [(p/2) !]^2}{[\Nn/2]^{p-1}},$$
where we allow for $U_p$ to be time dependent.
In this work we focus on a series of such $p/2$-body interacting SYK models
\begin{equation}
\Hh = \sum_{p} \Hh_{p}. \label{GenSYK}
\end{equation}
Specifically we will be interested in the \emph{many}-many body case, that is to say, the case where $p$ is large. However, the derivations which follow here are for the general case. We will introduce the details of the large $p$ case in Sec. \ref{ss:Many}.

By tuning a chemical potential, we are able to consider the system at arbitrary filling, encoded by the charge density
\begin{equation}
\Qq \equiv \frac{1}{\Nn}\sum_{k=1}^{\Nn} \ex{c^\dag_{k} c_k} -\frac{1}{2}, \label{charge}
\end{equation}
which is a conserved quantity. For instance, half filling corresponds to charge neutrality $\Qq =0$, for which we will find the same equations as in the Majorana case \cite{Eberlein2017}. 

We are interested in the non-equilibrium dynamics of \eqref{GenSYK} which, following \cite{Eberlein2017}, we will study in the Keldysh formalism \cite{Kamenev2011}.  In this framework one computes correlations, such as the Green's functions
\begin{equation}
\Gg(t_1,t_2) \equiv \frac{-1}{\Nn} \sum_{k=1}^\Nn \ex{ \Tt_\Cc c_k(t_1)c_k^\dag(t_2)}, \label{GreenCont}
\end{equation}
by considering a closed time contour $\Cc$. Here $\Tt_\Cc$ is the contour time ordering operator. From this definition, we note the Green's functions satisfy the conjugate relation 
\begin{equation}
\Gg(t_1,t_2)^* = \Gg(t_2,t_1), \label{conju}
\end{equation}
 which we will use at a later stage. These functions encode various statistics of the model, such as the density of states and charge density. Their time evolution is determined by the self energy $\Sigma$ via the Dyson equation $\Sigma(t_1,t_2) = \i\dot{\delta}_{\Cc}(t_1,t_2)-\Gg^{-1}(t_1,t_2) $. The SYK models are solvable in the sense that, in the thermodynamic limit, one can derive a closed form expression for the self energy in terms of the Green's functions. As an example, the $p/2$-body interacting SYK model has a self energy which is related to the Green's functions via \cite{Davison2017,Fu2018} 
 \begin{align}
 \Sigma_p(t_1,t_2) &= p 2 U_{p}^2 [-4\Gg(t_1,t_2)\Gg(t_2,t_1)]^{p/2-1} \Gg(t_1,t_2) \label{SelfEn}
 \end{align}
 to leading order in $1/\Nn$. For a sum of SYK models, such as the model we consider \eqref{GenSYK}, the self energies are simply additive $\Sigma(t_1,t_2) = \sum_{p} \Sigma_{p}(t_1,t_2)$ \cite{Maldacena2016Nov}.

\section{Real-time dynamics \label{Dynamics}}
We are particularly interested in the relaxation dynamics of the Green's functions \eqref{GreenCont} for Hamiltonians \eqref{GenSYK}, which take on a time dependence. To study this, it is convenient to work with time arguments defined along the real number line, instead of the closed time contour $\Cc$. 
In this real-time formalism, we  focus on the forward and backwards Green's functions, where $t_1$, $t_2$ are chosen to lie on different halves of $\Cc$. These forward and backwards Green's functions may be written explicitly as
\begin{align}
\Gg^{>}(t_1,t_2) &\equiv - \frac{1}{\Nn} \sum_{k=1}^\Nn \ex{c_k(t_1)c_k^\dag(t_2)}\\
\Gg^{<}(t_1,t_2) &\equiv \frac{1}{\Nn} \sum_{k=1}^\Nn \ex{c_k^\dag(t_2)c_k(t_1)},\label{GreenDefs}
\end{align}
respectively.
 Their equations of motion, the Kadanoff-Baym (KB) equations, are obtained from the Dyson equation by applying the Langreth rules \cite{Stefanucci2013}
\begin{equation}
\p_{t_1}\Gg^{\gtrless}(t_1,t_2) = \!\!\nint[t_1]{t_2}{t_3}\Sigma^{\gtrless}(t_1,t_3) \Gg^{A}(t_3,t_2) + I(t_1,t_2). \label{KBboth}
\end{equation}

Here we have defined the advanced Green's function, which for $t_3<t_2$ is
\begin{equation}
\Gg^{A}(t_3,t_2) = \Gg^{<}(t_3,t_2)-\Gg^{>}(t_3,t_2).
\end{equation}
In App. \ref{AppKB}, we elaborate on the process of writing of the KB equations in this form. The final term is the integral
\begin{align}
I(t_1,t_2)\! =\! \! \nint[-\infty]{t_1}{t_3} \Sigma^{>}(t_1,t_3) \Gg^{<}(t_3,t_2)-\Sigma^{<}(t_1,t_3)\Gg^{>}(t_3,t_2) \label{Idef}
\end{align}
which is the same for both forward and backwards Green's functions. As $t_2 \to t_1$ it is the only remaining term in the KB equations \eqref{KBboth}, since the first integral drops out. It is also at such equal times, that the left-hand side of \eqref{KBboth} may be related to the energy via the Galitskii–Migdal sum rule \cite{Stefanucci2013,galitskii1958application}
\begin{equation}
\lim_{t_2 \to t} \p_{t}\Gg^{<}(t,t_2) = -\i\sum_{p > 0} \frac{p}{2} E_p(t), \label{Gal}
\end{equation}
where $E_p(t) = \ex{\Hh_p(t)/\Nn}$. We provide a proof for this relation in App. \ref{WeightEnergy}. Comparing \eqref{Gal} to the right-hand side of \eqref{KBboth} yields the following relation to the energy terms 
\begin{equation}
I(t,t) = -\i\sum_{p > 0} \frac{p}{2} E_p(t), \label{IsumRules}
\end{equation}
which will be key in solving the equations.

\subsection{\emph{Many} many-body case \label{ss:Many}}

In this work we focus on $p = \kappa q$-body terms, for large $q$, which are particularly amenable to analytic calculation.  In the regime where $1/q$ is small, the Green's functions take the form $\Gg^{\gtrless}(t_1,t_2) \propto 1+g^{\gtrless}(t_1,t_2)/q +\Oo(1/q^2)$ \cite{Maldacena2016Nov,Davison2017}. Conveniently, by first rescaling the couplings as
\begin{equation}
U_{p} = \frac{J_{p}}{\sqrt{pq}}, \label{Jpdef}
\end{equation}
this structure is preserved even if we also have a competing kinetic term 
\begin{equation}
\Hh_{2} = \sum_{i=1}^N \sum_{j=1}^N X^{ i}_{j} c^{\dag}_{ i} c_{j},  \quad \overline{\,|X^{ i}_{j}|^{2}} =  \frac{1}{\Nn} \frac{J_{2}^2}{q}. \label{HamKin}
\end{equation}
With this, the general Hamiltonian we consider is of the form
\begin{equation}
\Hh(t) = \Hh_2(t) + \sum_{\kappa > 0} \Hh_{\kappa q}(t). \label{GenHam}
\end{equation}
Here we again mention that at this point, the couplings $J_{p}(t)$ still have a general time dependence.

In this work we opt to write the $1/q$ expansion in an exponential form,
\begin{equation}
\Gg^{\gtrless}(t_1,t_2) = \left[\Qq \mp \frac{1}{2}\right] e^{g^{\gtrless}(t_1,t_2)/q}, \label{Large_qG}
\end{equation}
which, together with higher order corrections, has been found to have larger overlap with the exact $q=4$ solution \cite{Tarnopolsky2019}.  This form will also aid in the interpretation of our results. For instance, linear correction in $g$, like $\i (t_1-t_2)$, may then be identified with a phase in the Green's functions, instead of a secular (diverging) term. We stress, however, that to leading order in $1/q$, the results for both choices are the same.

 For general $p$, we write the self-energy \eqref{SelfEn} as 
\begin{align*}
\Sigma^{\gtrless}_{p}(t_1,t_2) &= \frac{1}{q} \Ll_p^{\gtrless}(t_1,t_2)  \Gg^{\gtrless}(t_1,t_2),
\end{align*}
where we have defined 
\begin{equation}
\Ll_{p}^{\gtrless}(t_1,t_2) \equiv 2 J_{p}(t_1)J_{p}(t_2) [-4\Gg^{\gtrless}(t_1,t_2)\Gg^{\lessgtr}(t_2,t_1)]^{p/2-1}. \label{Ll}
\end{equation}

Since the total self energy $\Sigma^{\gtrless} = \Ll^{\gtrless} \Gg^{\gtrless}/q$ is a sum over the individual terms, we have
\begin{equation}
\Ll^{\gtrless}(t_1,t_2) = 2 J_{2}(t_1) J_{2}(t_2)+ \sum_{\kappa > 0} \Ll_{\kappa q}^{\gtrless}(t_1,t_2). \label{Lkin}
\end{equation}
where we have written the kinetic term's contribution, which corresponds to $p=2$, out explicitly.

Considering the definition \eqref{GreenDefs}, we note that at $t_1 = t_2 = 0$ the Green's function are equal to the charge density up to a constant $\Gg^{\gtrless}(0,0) = \Qq \mp 1/2$, which implies the boundary conditions $g^{\gtrless}(0,0) = 0$.

Inserting the large $q$ Green's function's expression \eqref{Large_qG} into the KB equations \eqref{KBboth}, the left-hand side simplifies to $ \Gg^{\gtrless}(t_1,t_2) \p_{t_1} g^{\gtrless}(t_1,t_2)/q$. Next, since the Green's functions are constant to leading order $\Gg^{\gtrless}(t_1,t_2) \sim \Gg^{\gtrless}(0,0)$, the advanced Green's function is given by $\Gg^{A}(t_3,t_2) \sim 1$. With this, dividing \eqref{KBboth} by $\Gg^{\gtrless}(0,0)/q$, we are left with
\begin{equation}
\p_{t_1} g^{\gtrless}(t_1,t_2) = \!\!\nint[t_1]{t_2}{t_3}\Ll^{\gtrless}(t_1,t_3) + \frac{qI(t_1,t_2)}{\Gg^{\gtrless}(0,0)}. \label{KBboth3}
\end{equation}

Using the definition \eqref{Idef} of $I$, the final term in \eqref{KBboth3}  may be written as
\begin{align*}
\frac{qI(t_1,t_2)}{\Gg^{\gtrless}(t_1,t_2)} &\sim 2\Gg^{\lessgtr}(0,0) \nint[-\infty]{t_1}{t_3} \frac{\Ll^{>}(t_1,t_3) - \Ll^<(t_1,t_3)}{2},
\end{align*}
which is remarkably independent of $t_2$, to leading order in $1/q$. As such, it must be the same expression as that at $t_2 = t_1$
\begin{equation}
\frac{q I(t_1,t_1)}{\Gg^\gtrless(t_1,t_1)} = 2\Gg^\lessgtr(0,0) \i \I(t_1), \label{GalMig2}
\end{equation}
where we have labeled the integral by $\i \alpha(t_1)$. Together with the Green's functions relation to charge density \eqref{Large_qG}, we are left with
\begin{equation}
\p_{t_1}g^{\gtrless}(t_1,t_2) = \nint[t_1]{t_2}{t_3} \Ll^{\gtrless}(t_1,t_3) + \Qq \i \alpha(t_1) \mp \frac{\i \alpha(t_1)}{2}. \label{KBboth2}
\end{equation}

The most apparent simplification, due to the $1/q$ expansion, is that integrands have lost their $t_2$ dependence.  This time argument only makes an appearance in the integral bound. Hence, by application of the fundamental theorem of calculus, one may obtain the second order differential equation
\begin{equation}
\p_{t_1} \p_{t_2} g^>(t_1,t_2) = \Ll^>(t_1,t_2). \label{LiouSum}
\end{equation}

Using the form of the Green's functions \eqref{Large_qG} we may express $\Ll_{\kappa q}^>(t_1,t_2)$, defined in \eqref{Ll}, as
\begin{align}
 2\Jj_\kappa(t_1) \Jj_\kappa(t_2) \exp\!\left[\kappa \frac{g^{>}(t_1,t_2) + g^{<}(t_2,t_1)}{2}\right]. \label{Ll2}
\end{align}
Here we have defined the \emph{effective} couplings
\begin{equation}
\Jj_{\kappa}(t) \equiv  [1-4\Qq^2]^{\kappa q/4-1/2} J_{p}(t). \label{couplingDef}
\end{equation}
At half filling $\Qq=0$, they are equal to $J_{p}(t)$.  For finite, non-zero, charge densities, we have $1-4\Qq^2<1$ leading to a suppression of the effective coupling
$$\Jj_{1}(t) \equiv  [1-4\Qq^2]^{q/4-1/2} J_{q}(t) \xrightarrow{q\to\infty} 0,$$
except for large $J_{q} \sim [1-4\Qq^2]^{-q/4}$. As such, to maintain non-trivial interactions away from charge neutrality, the coupling can be rescaled by this factor \cite{Davison2017}. For our results, however, we do not need to specify the scaling of $J_{q}$.

\section{Single SYK term \label{s:Single}}

Considering the relation between the integral $I$ and the weighted sum over these terms \eqref{IsumRules}, we note that $\alpha(t_1)$, defined in \eqref{GalMig2}, is given by 
\begin{equation}
(1-4\Qq^2)\I(t) = q E_{2}(t) + q^2 \sum_{\kappa >0} \kappa E_{\kappa q}(t). \label{GalMig}
\end{equation}
Note here that all terms in this series contribute to the same order. This is because of the scaling in \eqref{Jpdef} leading to the kinetic term scaling like $1/q$, while the interaction terms scale like $1/q^2$. For a system in equilibrium, the individual terms $E_p(t)$ are all constant, leading to constant \eqref{GalMig}. Otherwise, even for constant couplings, the weighted sum \eqref{GalMig} will generally not be constant, since the individual terms $\Hh_p(t)$ are not conserved. In contrast, an equally weighted sum would correspond to the conserved (for constant couplings) total energy.
With this we note the remarkable simplification which occurs in the case where we switch off all but for a single coupling in \eqref{GenHam} 
$$\Hh(t) = \begin{cases} 
\Hh_2(t) + \sum_{\kappa > 0} \Hh_{\kappa q}(t) & t<0\\
\Hh_p & t\ge 0
\end{cases},$$

In this case, the total interaction energy density is merely given by the single expectation value $E = \ex{\Hh_p/\Nn}$. Since this is a conserved quantity, we find, using the relation \eqref{GalMig} that $\alpha(t\ge 0)$ must also be a constant $\alpha_f$.

To ensure the applicability of our KB equations, we require the system to be in equilibrium in the asymptotic past. Note, however, that we have \emph{not} made any additional assumptions on the initial state $\varrho(0)$ of the system.   The key to our proof is \emph{only} that the final Hamiltonian consists of a single SYK term.  This may be accomplished via a quench, as was considered in \cite{Eberlein2017} for $\Hh_{q} + \Hh_{2q} \to \Hh_{q}$, via a ramp or any other time protocol, as shown in fig. \ref{fig:coupling}. 

With this, the KB equations \eqref{KBboth2} simplify to
\begin{align}
\p_{t_1} g^{>}(t_1,t_2)
&= \nint[t_1]{t_2}{t_3} \Ll^>(t_1,t_3) +\i\I_f+2\Qq \i\I_f  \label{KB>}\\ 
\p_{t_1} g^{<}(t_2,t_1) &= \nint[t_1]{t_2}{t_3} \Ll^>(t_1,t_3) +\i\I_f -2\Qq\i \I_f. \label{KB<}
\end{align}
Here we  have used the Green's function's conjugate relations \eqref{conju} $\Gg(t_1,t_2)^* = \Gg(t_2,t_1)$, which imply that $g^{\gtrless}(t_1,t_2)^* = g^{\gtrless}(t_2,t_1)$, to replace $\Ll^<(t_1,t_3)^* \to \Ll^>(t_1,t_3)$.
Equations \eqref{KB>} and \eqref{KB<} are remarkably similar, differing only by a constant. With the same boundary conditions, $g^{\gtrless}(0,0) = 0$, their solutions can thus only differ by a linear term 
\begin{align}
g^{>}(t_1,t_2) &= g(t_1,t_2) + 2\Qq \I_f\,\i (t_1-t_2) \label{newg>}\\
g^{<}(t_2,t_1) &= g(t_1,t_2) - 2\Qq \I_f\,\i (t_1-t_2). \label{newg<}
\end{align}
To derive the linear $t_2$ dependence, we have again used the Green's function's conjugate relations \eqref{conju}. From this we note that the Majorana relation $g^{>}(t_1,t_2) = g^{<}(t_2,t_1)$, found in \cite{Eberlein2017}, is reproduced as $\Qq \to 0$. Considering the Green's functions \eqref{Large_qG}, we note that such a linear result yields a phase
\begin{equation}
\Gg^{>}(t_1,t_2) \propto  \exp\left(\i \frac{2\Qq \I_f\, (t_1-t_2)}{q} +  g(t_1,t_2)/q\right). \label{massshift}
\end{equation}

This corresponds to a frequency independent shift in the self energy $\Sigma(\omega)$.  In the next section, we will show how this manifests itself as a shift in the chemical potential.

The addition of \eqref{newg>} and \eqref{newg<} yields the symmetric sum
\begin{equation}
g(t_1,t_2) \equiv \frac{g^{>}(t_1,t_2) + g^{<}(t_2,t_1)}{2}, \label{gasdef}
\end{equation}
previously encountered in \eqref{Ll2}. 

\subsection{Single interaction term}
We now consider the protocol where we are left only with a single interaction term,i.e., $\Hh(t>0) = \Hh_{q}$.
With this,  as was found for Majorana fermions \cite{Eberlein2017}, \eqref{LiouSum} reduces to the Liouville equation
\begin{equation}
\p_{t_1}\p_{t_2} g(t_1,t_2)  = 2\Jj^2 e^{g(t_1,t_2)}. \label{LiouvilleEq}
\end{equation}
This equation is formally similar to the corresponding equilibrium equation $\ddot{g}(t) \propto e^{g(t)}$. The key difference however is that we have two different time arguments, while in equilibrium only the relative time enters. 
For $g^*(t_1,t_2) = g(t_2,t_1)$, the solution of \eqref{LiouvilleEq} may be written in the form \cite{Tsutsumi1980Jul}
\begin{align}
e^{ g(t_1,t_2)} &= \frac{-\dot{u}(t_1)\dot{u}^*(t_2)}{ \Jj^2 [u(t_1)-u^*(t_2)]^2}. \label{eofg}
\end{align}
We would next like to find the most general, \emph{unique} solution \eqref{eofg}. We may find $u(t)$ by considering the equal time sum of \eqref{KB>} and \eqref{KB<} $$\lim_{t_2 \to t}\p_{t} g(t,t_2) = \i \I_f.$$ 
Substituting in the expression \eqref{eofg} for $g$ yields
\begin{equation}
\p_t \ln \dot{u}(t) - 2 \frac{\dot{u}(t)}{u(t)-u^*(t^+)} = \i \I_f. \label{diffh}
\end{equation}
Following \cite{Eberlein2017}, we make the ansatz 
\begin{equation}
u(t) = \frac{a \,e^{\i \pi v/2} e^{\sigma t} + \i b}{c\, e^{\i \pi v/2} e^{\sigma t} + \i d}, \quad v \in [-1,1] \label{exact2}
\end{equation}
which has $5$ independent real parameters, since the numerator and denominator are unique only up to an overall factor. Substituting \eqref{exact2} into \eqref{diffh} we are left with a constant $-\i\sigma \tan(\pi v/2)$ on the left. Hence, identifying 
\begin{equation}
\I_f = -\sigma \tan(\pi v/2) \label{boundalpha+}
\end{equation}
we see that the ansatz solves the equation. This leaves only four free parameters, which are fully determined by the two complex initial conditions, implying that this ansatz is a valid and uniquely determined solution. By the Picard-Lindel{\" o}f (Cauchy-Lipschitz) theorem this is also the only solution to the non-linear differential equation \eqref{diffh}, determined by the $2$ complex initial conditions $u(0)$ and $\dot{u}(0)$. 

Substituting this into \eqref{eofg} the correction $g(t_1,t_2)$, for $t_1,t_2\ge 0$,  takes on the unique and \emph{most} general form
\begin{equation}
e^{g(t_1,t_2)} = \frac{(\sigma/2)^2}{\Jj^2 \cos^2(\pi v/2-\sigma \i (t_1-t_2)/2)}, \quad \sigma \ge 0. \label{postQuench2}
\end{equation}
Note that \emph{all} solutions of \eqref{LiouvilleEq} \emph{only} depend on relative time $t_1-t_2$, when both times are larger than zero, because $\I_f$ is time independent. Remarkably \eqref{postQuench2} is completely independent of $a,b,c,d$, implying an SL$(2,\mathbb{C})$ invariance discussed in \cite{Eberlein2017}. Here we have assumed without loss of generality that $\sigma >0$, since $\sigma \to -\sigma$, is equivalent to taking $v \to -v$. 

\subsubsection{Comparison to equilibrium \label{sec:Eq}}

As noted previously, like in the equilibrium case, the solution \eqref{postQuench2}, only depends on time differences $g(t_1-t_2) \equiv g(t_1,t_2)$. It is also a periodic function satisfying $g(t) = g(-t-\i 2\pi v/\sigma)$. Such an equation is in fact a Kubo-Martin-Schwinger (KMS) relation for a system with inverse temperature  
$\beta_f \equiv 2\pi v/\sigma$. With this identification of the temperature, we have the expression
\begin{equation}
\sigma = \frac{2\pi v}{\beta_f}, \label{Lyap}
\end{equation}
where $\sigma$ is in fact the Lyapunov exponent of the system \cite{Maldacena2016Nov,Bhattacharya2017}. 
Inserting this into \eqref{postQuench2}, the \emph{correction} $g$ takes on the standard large $q$ thermal Green's function form \cite{Maldacena2016Nov}
\begin{equation} 
e^{g(t)/2} = \frac{\pi v}{\beta_f \Jj \cos(\pi v(1/2 - \i t/\beta_f))}. \label{gsol}
\end{equation}

It must also satisfy the same boundary condition $g(0)=0$, which yields the same closure relation
\begin{equation}
\beta_f\Jj = \frac{\pi v}{\cos(\pi v/2)}. \label{closure}
\end{equation} 
From this relation we are able to find $v$ as a function of $\beta_f \Jj$. Given a particular energy density \eqref{boundalpha+}, one is then able to find the corresponding temperature.

One should note, however,  that for the \emph{total} system to be considered in thermal equilibrium, it is the \emph{full} Green's functions that must satisfy the KMS relation. As such, we turn our attention to the full exact Green's functions, defined in \eqref{GreenDefs}. For $t>0$ we find that the forward and backward Green's functions are related via
\begin{equation}
\frac{\Gg^{<}(t+\i\beta_f)}{-\Gg^{>}(t)} = \frac{1+2\Qq }{ 1-2\Qq} e^{-\beta_f  2\Qq \I_f/q}. \label{KMSpost}
\end{equation}
For a standard KMS relation, the right-hand side is $1$, while in the presence of a chemical potential this changes to \cite{Sorokhaibam2020Jul}
\begin{equation}
\Gg^{<}_{\text{eq}}(t+\i\beta) = -e^{\beta \mu} \Gg^{>}_{\text{eq}}(t). \label{chemPotDef}
\end{equation} 
This is because, when considering real-time dynamics, the chemical potential term enters the state $\propto e^{-\beta [\Hh -\mu \Nn\hat{\Qq}]}$, but not the Hamiltonian.

As such, for $t\ge0$, the system can be \emph{immediately} identified as being in \emph{thermal equilibrium}, with a new chemical potential term
\begin{equation}
\mu_f(\Qq) = T_f\ln\left[\frac{1+2\Qq }{ 1-2\Qq}\right] - 2\Qq \I_f/q. \label{mu}
\end{equation}

\subsubsection{The final energy range \label{sec:PropState}}
As a final consistency check, we show that the energy densities of our solutions are always bounded by the lowest and highest eigenvalues of $\Hh_q/N$. We start by first writing the energy density in a simpler form by use of the closure relation $g(0) = 0$, meaning $\sigma = 2\Jj \cos(\pi v/2)$. Recalling the relation \eqref{boundalpha+} $\I_f = -\sigma \tan(\pi v/2)$, the energy density \eqref{GalMig} $q^2E = (1-4\Qq^2) \alpha_f$ may then be expressed as  
\begin{equation}
q^2E = - (1-4\Qq^2) 2\Jj \sin(\pi v/2). \label{energy}
\end{equation} 
For $v \in [-1,1]$, its range is then given by $$q^2 E \in [-2\Jj\, (1-4\Qq^2) , 2\Jj\,(1-4\Qq^2) ].$$  Here, in fact, the lower bound corresponds to the ground state energy density \cite{Davison2017}. Due to the symmetry of the SYK spectrum  over the zero axis \cite{Garcia-Garcia2017Sep}, the maximal energy is $-E_0$. As such, the allowable energies span the spectrum of the model.

In summary, given a state $\varrho(0)$, generated by a general protocol such as that shown in fig. \ref{fig:coupling}, we find the instantaneously thermal Green's function correction $g(t)$, given in \eqref{gsol}
\begin{equation} 
e^{g(t)/2} = \frac{\pi v}{\beta_f \Jj \cos(\pi v(1/2 - \i t/\beta_f))}.
\end{equation}

The system only has memory of two observables, namely the charge density 
$$\Qq \equiv \frac{1}{\Nn}\sum_{k=1}^{\Nn} \tr\{c^\dag_{k} c_k \varrho(0)\} -\frac{1}{2},$$
which is between $[-1/2,1/2]$, where $\Qq=0$ at half-filling, and the energy density
$$E_q \equiv \frac{1}{\Nn}\tr\{\Hh_q \varrho(0)\}.$$

Together they uniquely determine the final thermal Green's function. In particular, the constant $v$ is determined by both densities
$$\sin(\pi v/2) = \frac{q^2 E_q}{- (1-4\Qq^2) 2\Jj}.$$

The effective coupling $\Jj$, defined in \eqref{couplingDef}, is given in terms of the charge density, and coupling $J_{q}$.
The final temperature is then determined by $v$ from the closure relation \eqref{closure}
$$T_f = \Jj\frac{\cos(\pi v/2)}{\pi v}.$$

\subsection{Single kinetic term \label{Kinetic}}
Before concluding, we briefly discuss the case where all interactions are switched off in \eqref{GenHam}, leaving only the kinetic term \eqref{HamKin}. With this, \eqref{LiouSum} reduces to only the leading term in \eqref{Lkin}  
$\p_{t_1}\p_{t_2} g(t_1,t_2)  = 2 J_2^2$. This equation has a quadratic solution which, together with the boundary conditions $g(0,0) = 0$ and $\p_{t} g(t,t^+) = \i \alpha_f$, is given by $$g(t_1,t_2) = - J_2^2 (t_1-t_2) (t_1-t_2-\i \alpha_f/J_2^2).$$ We again note that this is only dependent on time differences. The total Green's function is then given by
\begin{equation}
\Gg^{>}(t) \propto  e^{\i  2\Qq \I_f t/q -  J_2^2 t (t-\i \alpha_f/J_2^2)/q }.
\end{equation}
It again satisfies a KMS relation $g(t) = g(-t+\i \alpha_f/J_2^{2})$, identifying the  inverse temperature as $$\beta_f = -\alpha_f/J_2^{2}.$$

As such, even in this case, we find instantaneous thermalization to leading order in $1/q$.

\section{Conclusion \label{conclusion}}

In this paper we extend previous results in Ref.~\cite{Eberlein2017} by studying the non-equilibrium dynamics of general Dirac fermion SYK models (\ref{GenHam})
\begin{equation}
\Hh(t) = \Hh_2(t) + \sum_{\kappa > 0} \Hh_{\kappa q}(t)
\label{eq_genH}
\end{equation}
in the $q\rightarrow\infty$ limit. Specifically, we were interested in the thermalization dynamics of a state~$\varrho$ at time $t=0$, which was generated from an equilibrium state of a time-independent Hamiltonian of form~\eqref{eq_genH} in the asymptotic past that is then acted upon by an arbitrary time-dependent $\Hh(t)$ for $t<0$. For $t\geq 0$ we made the key assumption that only a single term in~\eqref{eq_genH} remains. Under this assumption we could show analytically that the local Green's function is instantaneously in equilibrium. The only properties of the state~$\varrho(t=0)$ that determine the $t\geq 0$ equilibrium behavior are the energy density and the charge density. {\em In this sense a single $q\rightarrow\infty$ SYK-term is a perfect thermalizer for a large class of states $\varrho$.} Notice that it is unimportant whether one arrives at the single SYK-term via a quench or a more general time-dependent protocol, as shown in fig. \ref{fig:coupling}.

We were able to prove this result by making use of the conserved quantities, namely the $q$-body interaction energy and charge density, in combination with the Galitskii-Migdal sum rule. This forced the Green's function to be constant along the diagonal, leading to a differential equation with a unique solution depending on two initial conditions, which we identified as the energy density and the charge density at time $t=0$. This unique solution turns out to be just the thermal Green's function of the Hamiltonian for $t\geq 0$.

This work was funded by the Deutsche Forschungsgemeinschaft (DFG, German Research Foundation) - 217133147/SFB 1073, project~B03 and the Deutsche akademische Austauschdienst (DAAD, German Academic Exchange Service).
\appendix

\section{Weighted energy \label{WeightEnergy}}

In this section, we prove the Galitskii–Migdal sum rule, by considering the simplest case of $n-$body interactions, where $n$ is even. From the explicit form of the backwards Green's function
\begin{align*}
\i\p_{t}\Gg^{<}(t^+,t) \equiv& \frac{\i}{N}\sum_{k} \ex{\p_{t} c_k^\dag(t)c_k(t^+)}\\
=& \frac{1}{\Nn}\sum_{k} \ex{[c_k^\dag,\Hh] c_k}(t)
\end{align*}

Explicitly, for any even $n$-body interaction term, we have
\begin{align*}
\sum_{k} [c_k^\dag ,c_{1} \cdots c_{n}]c_k  &= \sum_{k} \sum_{\nu=1}^{n} (-1)^{\nu-1} c_{1} \cdots \{c_k^\dag,c_{\nu}\} \cdots c_{n} c_k\\
&= -\sum_{k} \sum_{\nu=1}^n c_{1} \cdots c_k \delta_{k,\nu} \cdots c_{n}\\
&= -n c_{1} \cdots c_{n}.
\end{align*}

Using the same identity for a string of operators $c_1^\dag \cdots c_n^\dag$, one would find $[c_k^\dag,c_1^\dag \cdots c_n^\dag] = 0$. As such, for some Hamiltonian 
\begin{equation}
\Hh = \sum_{\substack{1\le i_1< \cdots < i_{p/2}\le \Nn \\ 1\le j_1<\cdots< j_{p/2}\le\Nn}} X^{i_{1}\cdots  i_{n}}_{j_{1}\cdots j_{n}} c^{\dag}_{ i_1} \cdots c^{\dag}_{ i_{n}} c_{j_{n}} \cdots c_{j_1}
\end{equation}
\vspace*{2pt}
using the identity $[c_k^\dag,C^\dag C] = C^\dag [c_k^\dag,C] + [c_k^\dag,C^\dag] C$, the second term will vanish. As such the commutator evaluates to
\begin{align*}
[c_k^\dag ,\Hh] =& \sum_{\substack{1\le i_1< \cdots < i_{p/2}\le \Nn \\ 1\le j_1<\cdots< j_{p/2}\le\Nn}} X^{i_{1}\cdots  i_{n}}_{j_{1}\cdots j_{n}} [c_k^\dag, c^{\dag}_{ i_1} \cdots c^{\dag}_{ i_{n}}] c_{j_{n}} \cdots c_{j_1}. 
\end{align*}

From the above expression $\sum_{k} [c_k^\dag ,\Hh] c_k$ is given by
\begin{align*}
-n \sum_{\substack{1\le i_1< \cdots < i_{p/2}\le \Nn \\ 1\le j_1<\cdots< j_{p/2}\le\Nn}} X^{i_{1}\cdots  i_{n}}_{j_{1}\cdots j_{n}} c^{\dag}_{ i_1} \cdots c^{\dag}_{ i_{n}} c_{j_{n}} \cdots c_{j_1},
\end{align*}
which is the Galitskii–Migdal sum rule \cite{Stefanucci2013} for $n$-body interactions 
\begin{equation}
\i\p_{t}\Gg^{<}(t^+,t) = - \frac{n}{\Nn} \ex{\Hh(t)}.
\end{equation}

\section{Kadanoff-Baym equations \label{AppKB}}
Using the Langreth rule, the full Kadanoff-Baym (KB) equations are given by \cite{Stefanucci2013}
\begin{widetext}
\begin{align*}
\p_{t_1} \Gg^{\gtrless}(t_1,t_2) &= -\nint[t_0]{\infty}{t_3} \Sigma^{\gtrless}(t_1,t_3) \Gg^{A}(t_3,t_2) + \Sigma^{R}(t_1,t_3) \Gg^{\gtrless}(t_3,t_2) +\nint[t_0 -\i\beta]{t_0}{t_3}  \Sigma^{<}(t_1,t_3) \Gg^{>}(t_3,t_2)
\end{align*}
\end{widetext}
where we have defined the advanced/retarded functions 
$$\Gg^{A}(t_3,t_2) = \Theta(t_2-t_3) [\Gg^{<}(t_3,t_2)-\Gg^{>}(t_3,t_2)]$$
$$\Sigma^{R}(t_1,t_3) = - \Theta(t_1-t_3) [\Sigma^{<}(t_1,t_3)-\Sigma^{>}(t_1,t_3)].$$

Under the Bogoliubov principle, the assumption that initial correlations become irrelevant as $t_0 \to -\infty$ \cite{Semkat1999Feb,Pourfath2007Jul}, the imaginary part of the contour is ignored. The KB equations then take the form
$$\nint[-\infty]{t_2}{t_3}\Sigma^{\gtrless}(t_1,t_3) \Gg^{A}(t_3,t_2)+\nint[-\infty]{t_1}{t_3} \Sigma^{R}(t_1,t_3) \Gg^{\gtrless}(t_3,t_2)$$

This may be written as
\begin{align*}
\p_{t_1} \Gg^{\gtrless}(t_1,t_2) =& \nint[t_1]{t_2}{t_3}\Sigma^{\gtrless}(t_1,t_3) \Gg^{A}(t_3,t_2) + I(t_1,t_2)
\end{align*}
where we have pulled out part of the first integral and combined it with the second, yielding
\begin{align*}
I(t_1,t_2) = \nint[-\infty]{t_1}{t_3} &\bigg[\Sigma^{\gtrless}(t_1,t_3) [\Gg^{<}(t_3,t_2)-\Gg^{>}(t_3,t_2)]\\
&- [\Sigma^{<}(t_1,t_3)-\Sigma^{>}(t_1,t_3)] \Gg^{\gtrless}(t_3,t_2)\bigg].
\end{align*}
In both $\gtrless$ cases, this integral reduces to
$$I(t_1,t_2) = \nint[-\infty]{t_1}{t_3}[\Sigma^{>}(t_1,t_3) \Gg^{<}(t_3,t_2)-\Sigma^{<}(t_1,t_3)\Gg^{>}(t_3,t_2)].$$
\bibliography{../../References/Bib.bib}
\end{document}